\documentclass[runningheads]{llncs}
\usepackage{graphicx}
\usepackage{amsmath}
\usepackage{amssymb}
\usepackage{multirow}
\usepackage[misc,geometry]{ifsym} 
\makeatletter
\newcommand{\@chapapp}{\relax}%
\makeatother

\usepackage[title,toc,titletoc,header]{appendix}

\DeclareMathOperator*{\argmin}{arg\,min}

\begin{document}
\title{Analyzing the Quantum Annealing Approach for
Solving Linear Least Squares Problems}
\titlerunning{Analyzing QA for lin least squares}

\author{Ajinkya Borle\textsuperscript{(\Letter)} \and
Samuel J. Lomonaco}
\authorrunning{Borle and Lomonaco}
\institute{CSEE Department,University of Maryland Baltimore County, Baltimore MD 21250
\email{\{aborle1,lomonaco\}@umbc.edu}\\
}

\maketitle              

\begin{abstract}
 With the advent of quantum computers, researchers are exploring if quantum mechanics can be leveraged to solve important problems in ways that may provide advantages not possible with conventional or classical methods. A previous work by O\textsc{\char13}Malley and Vesselinov in 2016 briefly explored using a quantum annealing machine for solving linear least squares problems for real numbers. They suggested that it is best suited for binary and sparse versions of the problem. In our work, we propose a more compact way to represent variables using two's and one's complement on a quantum annealer. We then do an in-depth theoretical analysis of this approach, showing the conditions for which this method may be able to outperform the traditional classical methods for solving general linear least squares problems. Finally, based on our analysis and observations, we discuss potentially promising areas of further research where quantum annealing can be especially beneficial.

\end{abstract}
\section{Introduction}
Quantum computing opens up a new paradigm of approaching computational problems that may be able to provide advantages that classical (i.e.  conventional) computation cannot match. A specific subset of quantum computing is the quantum annealing meta-heuristic, which is aimed at optimization problems.

Quantum annealing is a hardware implementation of exploiting the effects of quantum mechanics in hopes to get as close to a global minimum of the objective function \cite{boixo2014evidence}.
One popular model of an optimization problem that quantum annealers are based upon is the Ising Model \cite{kadowaki1998quantum}. It can be written as:
\begin{align}
    F(h,J) = \sum_{a}h_a\sigma_a + \sum_{a<b}J_{ab}\sigma_{a}\sigma_{b}\label{eq:ising}
\end{align}
where $\sigma_{a} \in \{-1,1\}$ represents the qubit (quantum bit) spin and $h_a$ and $J_{ab}$ are the coefficients for the qubit spins and couplers respectively \cite{dorband2016stochastic}.
The quantum annealer's job is to return the set of values for $\sigma_{a}$s that would correspond to the smallest value of $F(h,J)$.

There have been various efforts by different organizations to make non Von Neumann architecture computers based on the Ising model such as D-wave Systems \cite{karimi2012investigating} and IARPA's QEO effort \cite{iarpa} that are attempting to make quantum annealers. Other similar efforts are focussed towards making `quantum like' optimizers for the Ising model, such as Fujistu's Digital Annealer chip \cite{aramon2018physics} and NTT's photonic quantum neural network \cite{honjo2018long}. The former is a quantum inspired classical annealer and the latter uses photonic qubits for doing its optimization. At the time of writing this document, D-wave computers are the most prominent Ising model based quantum annealers.

In order to solve a problem on a quantum annealer, the programmers first have to convert their problem for the Ising model. A lot of work has been done showing various types of problems running on D-wave machines \cite{adachi2015application,o2017nonnegative,o2015bayesian}.

In 2016, O'Malley and Vesselinov \cite{o2016toq} briefly explored using the D-wave Quantum Annealer for the linear least squares problem for binary and real numbers. In this paper, we shall study their approach in more detail. Section 2 is devoted to the necessary background and related work for our results. Section 3 is a review of the quantum annealing approach where we introduce one's and two's complement encoding for a quantum annealer. Section 4 deals with the runtime cost analysis and comparison where we define necessary conditions for expecting a speedup. Section 5 is dedicated to theoretical accuracy analysis. Based on our results, Section 6 is a discussion which lays out potentially promising future work. We finally conclude our paper with Section 7. The D-wave 2000Q and the experiments performed on them are elaborated upon in Appendices A and B respectively.

\section{Background and Related Work}
\subsection{Background}
Before we get started, we shall first lay out the terms and concepts we will use in the rest of the paper.
\paragraph{\textbf{Quantum Annealing:}}
The Quantum Annealing approach aims to employ quantum mechanical phenomena to traverse through the energy landscape of the Ising model to find the ground state configuration of $\sigma_a$ variables from Eqn(\ref{eq:ising}). The $\sigma_a$ variables are called as qubits spins in quantum annealing, essentially being quantum bits. 

The process begins with all qubits in equal quantum superposition: where all qubits are equally weighted to be either -1 or +1. After going through a quantum-mechanical evolution of the system, given enough time, the resultant state would be the ground state or the global minimum of the Ising objective function. During this process, the entanglement between qubits (in the form of couplings) along with quantum tunneling effects (to escape being stuck in configurations of local minima) plays a part in the search for the global minimum. A more detailed description can be found in the book by Tanaka et al. \cite{tanaka2017quantum}. For our purposes however, we will focus on the computational aspects related to quantum annealing: cost of preparing the problem for the annealer, cost of annealing and the accuracy of the answers. From an accuracy perspective, a quantum annealer is essentially trying to take samples of a Boltzmann distribution whose energy is the Ising objective function \cite{adachi2015application}
\begin{align}
    P(\sigma) &= \frac{1}{Z} e^{-F(h,J)}\label{eq:prob_boltzmann_ising}\\
   \text{where } Z &= \textit{exp}\big(\sum_{\{\sigma_{a}\}}\big[ \sum_{a}h_a\sigma_a + \sum_{a<b}J_{ab}\sigma_a\sigma_b \big] \big)\label{eq:prob_partition_ising}
\end{align}
Eqn(\ref{eq:prob_boltzmann_ising}) tells us that the qubit configuration of the global minimum would have the highest probability to be sampled.
The D-wave 2000Q is one such quantum annealer made by D-wave Systems. Its description is in Appendix A.
\paragraph{\textbf{Quadratic Unconstrained Binary Optimization (QUBO):}}These are minimization problems of the type
\begin{align}
    F'(v,w) = \sum_{a}v_{a}q_{a} + \sum_{a<b}w_{ab}q_{a}q_{b}\label{eq:qubo}
\end{align}
where $q_{a} \in \{0,1\}$ are the qubit variables returned by the machine after the minimization and $v_{a}$ and $w_{ab}$ are the coefficients for the qubits and the couplers respectively \cite{dorband2016stochastic}. 
The QUBO model is equivalent to the Ising model by the following relationship between $\sigma_{a}$ and $q_{a}$
\begin{align}
    \sigma_{a} &= 2q_{a} -1\\
    \text{and } F(h,J) &= F'(v,w) + \text{offset}
\end{align}
Since the offset value is a constant, the actual minimization is only done upon $F$ or $F'$. We use this model for the rest of the paper.
\paragraph{\textbf{Linear Least Squares:}}
Given a matrix $A \in \mathbb{R}^{m \times n}$, a column vector of variables $x \in \mathbb{R}^n$ and a column vector $b \in \mathbb{R}^m$ (Where $m>n$). The linear least squares problem is to find the $x$ that would minimize $\|Ax-b\|$ the most. In other words, it can be described as:
\begin{align}
    \argmin_x \|Ax - b\| \label{eq:LLS}
\end{align}
Various classical algorithms have been developed over the time in order to solve this problem. Some of the most prominent ones are (1)Normal Equations by Cholesky Factorization, (2) QR Factorization and the (3) SVD Method \cite{do2012numerically}.

\subsection{Related Work}
The technique of using quantum annealing for solving linear least squares problems for real numbers was created by O'Malley and Vesselinov \cite{o2016toq}. In that work, they discovered that because the time complexity is in the order of $O(mn^2)$, which is the same time complexity class as the methods mentioned above, the approach might be best suited for binary and sparse least squares problem. In a later work, O'Malley et al. applied binary linear least squares using quantum annealing for the purposes of nonnegative/binary matrix factorization \cite{o2017nonnegative}.

The problem of solving linear least squares has been well studied classically. Like mentioned above, the most prominent methods of solving the problems are Normal Equations (using Cholesky Factorization), QR Factorization and by Singular Value Decomposition \cite{do2012numerically}. But other works in recent years have tried to get a better time complexity for certain types of matrices, such as the work by Drineas et. al that presents a randomized algorithm in $O(mn\log n)$ for $(m>>n)$ \cite{drineas2011faster}. The iterative approximation techniques such as Picani and Wainwright's work \cite{pilanci2016iterative} of using sketch Hessian matrices to solve constrained and unconstrained least squares problems. But since the approach by O'Malley and Vesselinov is a direct approach to solve least squares, we shall focus on comparisons with the big 3 direct methods mentioned above.

Finally, it is important to note that algorithms exists in the  gate-based quantum computation model like the one by Wang \cite{wang2017quantum} that runs in poly$(\log(N),d,\kappa,1/\epsilon)$ where N is the size of data, $d$ is the number of adjustable parameters,$\kappa$ represents the condition number of $A$ and $\epsilon$ is the desired precision. However, the gate based quantum machines are at a more nascent stage of development compared to quantum annealing.

\section{Quantum Annealing for Linear Least Squares}
In order to solve Eqn(\ref{eq:LLS}), let us begin by writing out $Ax-b$
\begin{align}
Ax -b =
\begin{pmatrix}
A_{11} & A_{12} & ... & A_{1n}\\
A_{21} & A_{22} & ... & A_{2n}\\
\vdots & \vdots & \vdots & \vdots\\
A_{m1} & A_{m2} & ... & A_{mn}
\end{pmatrix}
\begin{pmatrix}
x_1\\
x_2\\
\vdots\\
x_n
\end{pmatrix}
-
\begin{pmatrix}
b_1\\
b_2\\
\vdots\\
b_m
\end{pmatrix}\\
Ax -b =
\begin{pmatrix}
A_{11}x_{1} + A_{12}x_{2} + ... + A_{1n}x_{n} - b_{1}\\
A_{21}x_{1} + A_{22}x_{2} + ... + A_{2n}x_{n} - b_{2}\\
\vdots\\
A_{m1}x_{1} + A_{m2}x_{2} + ... + A_{mn}x_{n} - b_{m}
\end{pmatrix} \label{eq:Ax-b}
\end{align}
Taking the 2 norm square of the resultant vector of Eqn(\ref{eq:Ax-b}), we get
\begin{align}
    \|Ax-b\|_{2}^{2} = \sum_{i=1}^{m}(| A_{i1}x_{1} + A_{i2}x_{2} +...+ A_{in}x_{n} - b_{i}|)^2
\end{align}
Because we are dealing with real numbers here, $(|.|)^2 = (.)^2$
\begin{align}
    \|Ax-b\|_{2}^{2} = \sum_{i=1}^{m}(A_{i1}x_{1} + A_{i2}x_{2} +...+ A_{in}x_{n} - b_{i})^2 \label{eq:2normsq}
\end{align}
Now if we were solving binary least squares \cite{o2016toq,o2017nonnegative} then each $x_{j}$ would be represented by the qubit $q_j$. The coefficients in Eqn(\ref{eq:qubo}) are found by expanding Eqn(\ref{eq:2normsq}) to be
\begin{align}
    v_{j} = \sum_{i}A_{ij}(A_{ij}-2b_{i})\label{eq:vbin}\\
    w_{jk} = 2\sum_{i}A_{ij}A_{ik}\label{eq:wbin}
\end{align}
But for solving the general version of the least squares problem we need to represent $x_j$, which is a real number, in its equivalent radix 2 approximation by using multiple qubits.
Let $\Theta$ be the set of powers of 2 we use to represent every $x_{j}$, defined as
\begin{align}
    \Theta = \{2^{l}:l \in [o,p] \land l,o,p\in\mathbb{Z}\}
\end{align}
Here, it is assumed that $l$ represents contiguous values from the interval of $[o,p$]. The values of $o$ and $p$ are the user defined lower and upper limits of the interval. In the work by O'Malley and Vesselinov \cite{o2016toq}, the radix 2 representation of $x_j$ is given by
\begin{align}
    x_{j} \approx \sum_{\theta \in \Theta} \theta q_{j\theta} \label{eq:radix2_1}
\end{align}
But this would mean that only approximations of positive real numbers can be done, so we need to introduce another set of qubits $q_{j}^{*}$, to represent negative real numbers
\begin{align}
    x_{j} \approx \sum_{\theta \in \Theta} \theta q_{j\theta} + \sum_{\theta \in \Theta} -(\theta q^{*}_{j \theta})\label{eq:radix2_2}
\end{align}
 Which means that representing a (fixed point approximation of) real number that can be either positive or negative would require $2|\Theta|$ number of qubits. 

However, we can greatly reduce the amount of qubits to be used in Eqn(\ref{eq:radix2_2}) by introducing a sign bit $q_{j\emptyset}$
\begin{align}
    x_{j} &\approx \vartheta q_{j\emptyset} + \sum_{\theta \in \Theta} \theta q_{j\theta}\label{eq:radix2_2comp}\\
    \text{where } \vartheta &= 
    \begin{cases}
    -2^{p+1}, & \text{for two's complement}\\
    -2^{p+1}+2^{o}, & \text{for one's complement}
    \end{cases}\label{eq:radix2_1comp}
\end{align}
Where $p$ and $o$ are the upper and lower limits of the exponents used for powers of 2 present in $\Theta$. In other words, Eqn(\ref{eq:radix2_2comp}) represents an approximation of a real number in one's or two's complement binary.
Combining Eqn(\ref{eq:2normsq}) and Eqn(\ref{eq:radix2_2comp}), we get
\begin{align}
\begin{split}
    \|Ax-b\|_{2}^{2} =
    \sum_{i=1}^{m}(A_{i1}(\vartheta q_{1\emptyset} + \sum_{\theta \in \Theta} \theta q_{1\theta})\\
    + A_{i2}(\vartheta q_{2\emptyset} + \sum_{\theta \in \Theta} \theta q_{2\theta}) +...+ A_{in}(\vartheta q_{n\emptyset} + \sum_{\theta \in \Theta} \theta q_{n\theta}) - b_{i})^2
\end{split} \label{eq:quadform}
\end{align}
Which means that the $v$ and $w$ coefficients of the qubits for the general version of the least squares problem would be
\begin{align}
    v_{js} &= \sum_{i} s A_{ij} (s A_{ij} - 2b_{i})\label{eq:v}\\
    w_{jskt} &= 2st \sum_{i}A_{ij}A_{ik} \label{eq:w}
\end{align}
where $s,t\in \vartheta \cup \Theta$
\section{Cost analysis and comparison}
In order to analyze the cost incurred using a quantum annealer to solve a problem, one good way is by combining together the (1) time required to prepare the problem (so that it is in the QUBO/Ising model that the machine would understand)  and (2) the runtime of the problem on the machine.

We can calculate the first part of the cost concretely. But the second part of the cost depends heavily upon user parameters and heuristics to gauge how long should the machine run and/or how many runs of the problem should be done. Nonetheless, we can set some conditions that must hold true if any speedup is to be observed using a quantum annealer for this problem.
\subsection{Cost of preparing the problem}
As mentioned in O'Malley and Vesselinov \cite{o2016toq} the complexity class of preparing a QUBO problem from $A$,$x$ and $b$ is $O(mn^2)$, which is the same as all the other prominent classical methods to solve linear least squares \cite{do2012numerically}. However, for numerical methods, it is also important to analyze the floating point operation cost as they grow with the data. This is because methods in the same time complexity class may be comparatively faster or slower.

So starting with Eqn(\ref{eq:v}), we assume that the values of the set $\vartheta \cup \Theta$ are preprocessed. Matrix $A$ has $m$ rows and $n$ columns, the variable vector $x$ is of the length n. Let $c= |\Theta| + 1$. To calculate the expression $s A_{ij} (s A_{ij} - 2b_{i})$, we can compute $2b_{i}$ for $m$ rows first and use the results to help in all the future computations. After that, we see that it takes 3 flops to process the expression for 1 qubit of the radix 2 representation of a variable, per row. This expression has to be calculated for $n$ variables each requiring $c$ qubits for the radix 2 representation, over m rows. That is: it would take $3cmn$ operations. On top of that, we need to sum up the resulting elements over all the $m$ rows, which requires an additional $cmn$ operations. Hence we have
\begin{align}
    \text{Total cost of computing } v_{js} = 4cmn + m \label{eq:cost_v}
\end{align}
Now for the operation costs associated with terms processed in Eqn(\ref{eq:w}). Let us consider a particular subset of those terms:
\begin{align}
    w_{j1k1} = 2\sum_{i}A_{ij}A_{ik}\label{eq:w0}
\end{align}
By computing Eqn(\ref{eq:w0}) first, we create a template for all the other $w_{jskt}$ variables and also reduce the computation cost.Each $A_{ij}A_{ik}$ operation is 1 flop. There are $\binom{n}{2}$ pairs of $(j,k)$ for $j < k$, But we also need to consider pair interactions for qubits when $j=k$. Hence we have $\binom{n}{2} + n$ operations, which comes out to $0.5(n^2 + n)$. Furthermore, these $w$ coefficients are computed for $m$ rows with $m$ summations, which brings the total up to: $m(n^2+n)$.
After this, we need to multiply $2$ to all the resultant $0.5(n^2+n)$ variables. Making the total cost:
\begin{align}
    \text{Cost of all $w_{j1k1}$} =m(n^2+n) +0.5(n^2+n) \label{eq:cost_w0}
\end{align}
Now we can use $w_{j1k1}$ for the next step. Without loss in generality, we assume that $\forall s,t \in \vartheta \cup \Theta$, $s \times t$ is preprocessed. This would mean that we would have $\binom{c}{2} + c$ qubit to qubit interactions for each pair of variables in $x$. From the previous step, we know that we'll have to do this for $\binom{n}{2} + n$ variables. Which means that the final part of the cost for $w_{jskt}$ is $0.25(c^2+c)(n^2+n)$.
Summing up all the costs, we get the total cost to prepare the entire QUBO problem: 
\begin{align}
    \text{Cost of tot. prep} = mn^2 + mn(4c+1) + 0.25(n^2+n)(c^2+c+2) + m
\end{align}
\subsection{Cost of executing the problem} \label{sec:cost_exec}
Let $\tau$ be the cost of executing the QUBO form of a given problem. It can be expressed as
\begin{align}
    \tau = a_t r
\end{align}
Where $a_t$ is the anneal time per run and $r$ is the number of runs or samples.
However, for ease of analysis, we need to interpret $\tau$ in a way where we can study the runtime in terms of the data itself. The nature of the Ising/Qubo problem is such that we need O(exp($\gamma N^{\alpha}$)) time classically to get the ground state with a probability of 1. As stated in Boxio et al.  \cite{boixo2014evidence}, we don't yet know what's going to be the actual time complexity class under quantum annealing for the Ising problem, but a safe bet is that it won't  reduce the complexity of the Ising to a polynomial one, only the values of $\alpha$ and $\gamma$ would be reduced. But because quantum annealing is a metaheuristic for finding the ground state, we needn't necessarily run it for O(exp($\gamma N^{\alpha}$)). We make the following assumption:
\begin{align}
    \tau^* = poly(cn) \label{eq:tau_prop}\\
    \text{and deg}(poly(cn)) = \beta \label{eq:beta}
\end{align}
Here, $\tau^{*}$ represents the combined operations required for annealing as well as post-processing on the returned samples. The assumption is that $\tau^{*}$ is a polynomial in $cn$ with $\beta$ as its degree. From Eqn(\ref{eq:quadform}), we know that $m$ (number of rows of the matrix $A$) doesn't play a role in deciding the size of the problem embedded inside the quantum annealer. 

The reason for this assumption is the fact that linear least squares is a convex optimization problem. Thus, even if we don't get the global minimum solution, the hope is to get samples on the convex energy landscape that are close to the global minimum (Based on the observations in \cite{dorband2018method}). Using those samples, and exploiting the convex nature of the energy landscape, the conjecture is that there exists a polynomial time post-processing technique with $\beta<3$ by which  we can converge to the global minimum very quickly. We shall see in section \ref{sec:comparison} why it is important for $\beta<3$ for any speed improvement over the standard classical methods. Just as in Dorband's MQC technique  \cite{dorband2018method} for generalized Ising landscapes, we hope that such a technique would be able to intelligently use the resultant samples. The difference being that we have the added advantage of convexity in our specific problem.

\subsection{Cost comparison with classical methods} \label{sec:comparison}
In the following table, we compare the costs of the most popular classical methods for finding linear least squares \cite{do2012numerically} and the quantum annealing approach.
\begin{table}[!ht]
\centering
\caption{Comparison of the classical methods and QA}
\label{table:1}
 \begin{tabular}{|c|c|}
 \hline
 \textbf{Method for Least Squares} & \textbf{Operational Cost} \\ [0.5ex] 
 \hline
 Normal Equations & $mn^2 + n^3/3$\\ 
 \hline
 QR Factorization & $2mn^2 -2n^3/3$ \\
 \hline
 SVD & $2mn^2 + 11n^3$ \\
 \hline
 Quantum Annealing & \begin{tabular}{@{}c@{}}$mn^2 + mn(4c+1) +  poly(cn)$\\ + $0.25(n^2+n)(c^2+c+2) + m$\end{tabular}\\
 \hline
\end{tabular}
\end{table}

When it comes to theoretical runtime analysis, because $c$ doesn't necessarily grow in direct proportion to the number of rows $m$ or columns $n$, we shall consider $c$ to be a constant for our analysis.

From Table \ref{table:1}, Let us define $Cost_{NE}$, $Cost_{QR}$, $Cost_{SVD}$ and $Cost_{QA}$ as the costs for the methods of finding least squares solution by Normal Equations, QR Factorization, Singular Value Decomposition and Quantum Annealing (QA) respectively.
\begin{align}
    Cost_{NE} &= mn^2 + n^3/3\label{eq:ne}\\
    Cost_{QR} &= 2mn^2 -2n^3/3\label{eq:qr}\\
    Cost_{SVD} &= 2mn^2 + 11n^3\label{eq:svd}\\
    Cost_{QA} &= mn^2 + mn(4c+1) + poly(cn) + 0.25(n^2+n)(c^2+c+2)+m \label{eq:qa}
\end{align}
The degree of Eqn(\ref{eq:ne}),Eqn(\ref{eq:qr}),Eqn(\ref{eq:svd}) and Eqn(\ref{eq:qa}) is 3.  Since $m>n$, we can assess that $mn^2 > n^3$

The next thing we need to do is to define a range for $\beta$ (degree of $pol(cn)$) in such a way that Eqn(\ref{eq:qa}) will be competitive with the other methods. This is another assumption upon which a speedup is conditional.
\begin{align}
    \boxed{0 < \beta < 3}\label{eq:beta_bound}
\end{align}
This is done so that we do not have another term of degree 3.
Now, we turn our attention to the terms of the type $kmn^2$ where $k$ is the coefficient, we can see that the cost of the QA method is lesser than QR Factorization and SVD by a factor of 2. However, we still have to deal with the cost of the Normal Equations Method.
Let us define $\Delta Cost_{NE}$ and $\Delta Cost_{QA}$ as
\begin{align}
    \Delta Cost_{NE} &= Cost_{NE} - mn^2 = n^3/3\label{eq:ne_delta}\\
    \begin{split}
    \Delta Cost_{QA} &= Cost_{QA} - mn^2 = mn(4c+1) + poly(cn)\\ &\qquad \qquad \qquad \qquad + 0.25(n^2+n)(c^2+c+2)+m\label{eq:qa_delta}
    \end{split}
\end{align}
The degree of $\Delta Cost_{NE}$ is 3 while that of $\Delta Cost_{QA}$ is $<3$. For simplicity (and without loss of generality) we consider $mn(4c+1)$ and can ignore all the other similar and lower degree terms. The reason we can do this is because those terms grow comparatively slower than $n^3/3$, but the relationship between $n^3/3$ and $mn(4c+1)$ has to be clearly defined. Thus our simplified cost difference for the QA method is $\Delta Cost_{QA}^{*}$
\begin{align}
    \Delta Cost_{QA}^{*} = (4c+1)mn
\end{align}
We need to analyze the case where the quantum annealing method is more cost effective, i.e 
\begin{align}
    \Delta Cost_{QA}^{*} &< \Delta Cost_{NE}\\
    \text{or, } mn(4c+1) &< n^3/3\label{eq:cost_anal}
\end{align}
For this comparison, we need to define $m$ in terms of $n$
\begin{align}
    m = \lambda n\label{eq:lamdba} 
\end{align}
Using Eqn(\ref{eq:lamdba}) in Eqn(\ref{eq:cost_anal}), we get
\begin{align}
    \lambda n^2(4c+1) &< \frac{n^3}{3}\\
    \text{or, } \lambda (4c+1) &< \frac{n}{3}\\
    \text{or, } \lambda &< \frac{n}{3(4c+1)}\label{eq:lambdabound1}
\end{align}
Combining Eqn(\ref{eq:lambdabound1}) with the fact that $m>n$, we get :
\begin{align}
    \boxed{1 < \lambda < \frac{n}{3(4c+1)}} \label{eq:lamdabound2}\\
    \text{given }\frac{n}{3(4c+1)} > 1 \label{eq:lambdacond}
\end{align}
Thus, the Quantum Annealing method being faster than the Normal Equations method is for when Eqn(\ref{eq:beta_bound},\ref{eq:lamdabound2},\ref{eq:lambdacond}) holds true and is conditional on the conjectures described in Eqn(\ref{eq:tau_prop},\ref{eq:beta},\ref{eq:beta_bound}).

The above condition makes our speed advantage very limited to a small number of cases. However, it is important to note that the Normal Equations method is known to be numerically unstable due to the $A^TA$ operation involved \cite{do2012numerically}. 
The quantum annealing approach does not seem to have such types of calculations that would make it numerically unstable to the extent of Normal Equations method (because of the condition number of $A$), assuming the precision of qubit and coupler coefficients is not an issue. Thus for most practical cases, it competes with the QR Factorization method rather than the Normal Equations method.
\section{Accuracy analysis}
Because quantum annealing is a physical metaheuristic, it is important to analyze the quality of the results obtained from it. The results from our experiments are in Appendix B. We can define the probability of getting the global minimum configuration of qubits in the QUBO form by using Eqn(\ref{eq:prob_boltzmann_ising},\ref{eq:prob_partition_ising})
\begin{align}
    P(q) &= \frac{1}{Z'} e^{-F'(v,w)}\label{eq:prob_boltzmann}\\
   \text{where } Z' &= exp\big(\sum_{\{q_{a}\}}\big[ \sum_{a}v_aq_a + \sum_{a<b}w_{ab}q_aq_b \big] \big)
\end{align}
Which means that the set of solutions corresponding to the global minimum $\hat{Q}$ have the highest probability of all the possible solution states.
A problem arises when we need to use more qubits for better precision. This would mean that the set of approximate solutions $Q'$, would also increase as a result. The  net result would be that $\sum_{\hat{q}\in\hat{Q}}P(\hat{q}) < \sum_{q'\in Q'}P(q')$, which means that as the number of qubits used for precision increases, it would be harder to get the best solution directly from the machine. But like discussed in Section \ref{sec:cost_exec}, if the conjecture for a polynomial time post-processing technique with degree$<3$ holds true, we should be able to get the best possible answer, in a competitive amount of time, by using the results of a quantum annealer.

Another area of potential problems is the fact that quantum annealing happens on the actual physical qubits and its connectivity graph (see Appendix A for more information). This means that the energy landscape for the physical qubit graph is bigger than the one for the logical qubit graph.
This problem should be alleviated to a degree when and if the next generation of quantum annealers have a more dense connectivity between their physical qubits.
\section{Discussion and Future Work}
Based on our theoretical and experimental results (see Appendix B), we can see that there are potential advantages as well as drawbacks to this approach. Our work outlines the need of a polynomial time post-processing technique for convex problems (with degree$<3$),only then will this approach have any runtime advantage. Whether such a post-processing technique exists is an interesting open research problem. But based on our work, we can comment on few areas where quantum annealing may have a potential advantage. We have affirmed the conjecture that machines like the D-wave find good solutions (that may not be optimal) in a small amount of time \cite{o2017nonnegative} (Appendix B).

It may be useful to use quantum annealing for least squares-like problems, i.e. problems that require us to minimize $\|Ax-b\|$, but are time constrained in nature. Two such problems are: (i) The Sparse Approximate Inverse \cite{grote1997parallel} (SPAI) type preconditioners used in solving linear equations and (ii) the Anderson acceleration method for iterative fixed point methods \cite{walker2011anderson}. Both of these methods require approximate least squares solutions, but under time constraints. It would be interesting to see if quantum annealing can be potentially useful there.

Another area of work could be to use quantum annealing within the latest iterative techniques for least squares approximation itself. Sketch based techniques like the Hessian sketch by Pilanci and Wainwright \cite{pilanci2016iterative} may be able to use quantum annealing as a subroutine.

Finally, just like O'Malley and Vesselinov mentioned in their papers  \cite{o2016toq,o2017nonnegative}, quantum annealing also has potential in specific areas like the Binary \cite{tsakonas2011robust} and box-constrained integer least squares \cite{chang2008solving} where classical methods struggle.

\section{Concluding Remarks}
In this paper, we did an in-depth theoretical analysis of the quantum annealing approach to solve linear least squares problems. We proposed a one's complement and two's complement representation of the variables in qubits. We then showed that the actual annealing time does not depend on the number of rows of the matrix, just the number of columns/variables and number of qubits required to represent them. We outlined conditions for which quantum annealing will have a speed advantage over the prominent classical methods to find least squares. An accuracy analysis shows how as precision bits are added, it is harder to get the 'best' least square answer, unless any post-processing is applied. Finally, we outline possible areas of interesting research work that may hold promise.

\subsubsection{Acknowledgement} The authors would like to thank Daniel O'Malley from LANL for his feedback. A special thanks to John Dorband, whose suggestions inspired the development the one's/two's complement encoding. Finally, the authors would like to thank Milton Halem of UMBC and D-wave Systems for providing access to their machines.
%
%
%
\bibliographystyle{splncs04}
\bibliography{references}
\begin{appendices}
\renewcommand{\thesection}{\appendixname~\Alph{section}}
\section{The D-wave 2000Q}
\begin{figure}
\includegraphics[scale=0.22]{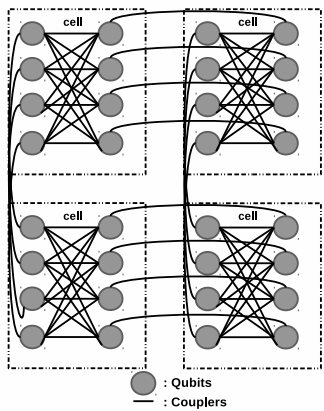}
\centering
\caption{Subsection of the Chimera Graph of the D-wave 2000Q} \label{fig:chimera}
\end{figure}

The D-wave 2000Q is a quantum annealer by D-wave Systems. At the time of writing, it is the only commercially available quantum annealer in the market. The 2048 qubits of its chip are connected in what is called a chimera graph. As represented in Figure \ref{fig:chimera}, a group of 8 qubits are grouped together in a cell in a bipartite manner. Each cell is connected to the cell below and the cell across as described in the figure. The chimera graph of this machine has 16 by 16 arrangement of cells, making it 2048 qubits.

We can see that the qubit connectivity is rather sparse. Here, we introduce the concept of logical qubits and physical qubits. Physical qubits are the ones on the machine connected in the chimera graph and the logical qubits are the qubits that describe the nature of our application problem, which may have a different connectivity graph. Thus, in order to run a problem on the D-wave, the logical qubit graph (in the ising model) is mapped or embedded on to the chimera graph by employing more than one physical qubit to represent one logical qubit. This is called qubit chaining.This is done by using particular penalties and rewards on top of the coefficients so that qubits of a chain that behave well (if one of them is +1 then all should be +1 and vice versa) are rewarded and are penalized if they don't.
After a problem has been embedded onto the physical qubits, the actual annealing process happens. Each run of the machine (called an anneal cycle) takes a fixed amount of time to execute that can be set between $1\mu$s to $2000\mu$s for the D-wave 2000Q. Because a quantum annealer is a probabilistic machine, we need to have a number of runs and tally up the results. We then take the result that is able to minimize the objective function the most, unembed it back to the logical qubit graph. The final configuration of the logical qubits is our answer.

For the D-wave 2000Q, the value that can be given to the coefficients is actually bounded as $h_a \in [-2,2]$ and $J_{ab} \in [-1,1]$. The precision of the coefficient is also low, the community usually works in the range of 4 to 5 bits, but the actual precision is somewhere about 9 bits\cite{dorband2018extending}. From a programmer's perspective, the API subroutines takes care of any arbitary floating point number by normalizing it down to the $h_a$ and $J_{ab}$ bounds, within the above mentioned precision of course.
\section{Experiments on the D-wave 2000Q}
Here we shall test the D-wave 2000Q for some Linear Least squares problems and compare the results with the classical solution in terms of accuracy. Due to the limited size of the current machine, our experiments use small datasets and therefore, are not able to provide a runtime advantage. The main aim is to assess the quality of the solutions that are returned by the D-wave without post-processing. Our language of choice is MATLAB\texttrademark.
\subsubsection{Experiment 1}Our first experiment compares the effectiveness of the solution obtained classically against the solution obtained by a D-wave machine. For the D-wave implementation, we use both types of encoding discussed in this paper: the basic encoding given in Eqn(\ref{eq:radix2_2}) and the ones' complement encoding given in Eqn(\ref{eq:radix2_2comp},\ref{eq:radix2_1comp}). The reason we use one's complement over two's complement is because of the limited precision for the coefficients available on the current generation of D-wave hardware.  As a standard practice\cite{o2016toq}, when it comes to the  D-wave machine, we are limiting the radix-2 approximation of each variable in $x$ to just 4 bits. Which means that we are limited to a vector $x$ with a length of 8 due to the basic encoding technique (which would require 64 fully connected logical qubits). We use the following code to generate the matrix $A$,$b$ and $x$.
\begin{verbatim}
rng(i)
A = rand(100,8)
A = round(A,3)
b = rand(100,1)
b = round(b,3)
x = A\b
\end{verbatim}
We do this for the range $\{i\in \mathbb{Z}|4 \leq i \leq 7 \}$ to generate 4 sets of $A$,$b$ and their least squares solution $x$. Because our primary aim is to see how well can the D-wave do given the current limitations, we first analyze the classical solutions obtained to see what contiguous powers of two can best capture the actual classical solution. We find that the absolute value of most variables in the classical solution lies between $2^{-2}$ and $2^{-5}$. Thus we use $\Theta = \{2^{-2},2^{-3},2^{-4},2^{-5}\}$. 

For testing purposes, we use an anneal time of $50 \mu$s and $10000$ samples for each problem. We can probably use a lower anneal time, but our choice of anneal time is aimed at providing a conclusive look at the quality of solutions returned by the D-wave while being still relatively low (the anneal time can go all the way up to $2ms$  on the D-wave 2000Q). The \textit{Minimize\_energy} technique is used to resolve broken chains. We run the \textit{find\_embedding} subroutine of the D-wave API 30 times and select the smallest embedding.
\begin{itemize}
    \item Basic encoding: physical qubits=1597,max qubit chain length=32
    \item One's complement: physical qubits=612,max qubit chain length=20
\end{itemize}
\begin{table}[!ht]
\centering
\caption{Experiment 1 (x has 8 variables)}
\label{table:2}
 \begin{tabular}{|c|c|c|c|c|}
 \hline
 \multirow{2}{*}{\textbf{Table No.}} & \multirow{2}{*}{\textbf{Rng Seed}} & \multicolumn{3}{|c|}{\textbf{Value of $\|Ax-b\|_{2}$}}\\
 \cline{3-5} &
 & \textbf{Classical Solution} & \textbf{D-wave (basic)} & \textbf{D-wave (1's comp)}\\
 \hline
 1 & 4 & 2.9873 & 2.9938 & 3.0014\\
 \hline
 2 & 5 & 2.9707 & 2.9751 & 2.9931\\
\hline
3 & 6 & 2.8 & 2.8054 & 2.8558\\
\hline
4 & 7 & 2.8528 & 2.8595 & 2.8887\\
\hline
\end{tabular}
\end{table}
In the results, we can see that the D-wave 2000Q can't arrive at the least square answer of the classical method's solution. Part of that can be explained by the limited amount of numbers that can be represented using 4 bits. We also see that using the basic encoding format, we can get a more effective answer than when the one's complement encoding is used. This can be attributed to the fact that in the basic encoding, a lot of the numbers within the 4 bit range can be represented in more than one way (because we have separate qubits for positive and negative bits). Which essentially means that the energy landscape of the basic encoding contains a lot of local and potentially global minimum states, all because of the redundancy in representation.The one's complement encoding is more compact and we can see that its energy landscape would have far fewer local minimas (and only one global minimum in most cases of linear least squares). But it does affirm the conjecture that the D-wave is able to find fast solutions that are good, but may not be optimal\cite{o2017nonnegative}.
\subsubsection{Experiment 2}For our second experiment, we generate $A \in \mathbb{R}^{100 \times 12}$, $x\in\mathbb{R}^{12\times1}$ and $b\in\mathbb{R}^{100\times1}$ to compare classical solutions with one's complement encoded results (the basic encoding can't do an $x$ vector of length above 8). The data is created in the same way as the previous experiment with the $x$ vector of  length 12 instead of 8. All the other parameters are kept the same. The best embedding we got (after 30 runs of \textit{find\_embedding} subroutine) has a total number of 1357 physical qubits with a 30 qubit max chain length.

\begin{table}[!ht]
\centering
\caption{Experiment 2 (x has 12 variables)}
\label{table:3}
 \begin{tabular}{|c|c|c|c|}
 \hline
 \multirow{2}{*}{\textbf{Table No.}} & \multirow{2}{*}{\textbf{Rng Seed}} & \multicolumn{2}{|c|}{\textbf{Value of $\|Ax-b\|_{2}$}}\\
 \cline{3-4} &
 & \textbf{Classical Solution} & \textbf{D-wave (1's comp)}\\
 \hline
 1 & 4 & 2.5366 & 2.5719 \\
 \hline
 2 & 5 & 2.6307 & 2.6773 \\
\hline
3 & 6 & 2.7942 & 2.8096 \\
\hline
4 & 7 & 2.9325 & 2.945 \\
\hline
\end{tabular}
\end{table}
We see similar results for this experiment like the one before it. But further experimental research needs to be done. Like mentioned in O'Malley and Vesselinov's work, we are limited by a lot of factors, some of them being
\begin{enumerate}
    \item Low precision for the coefficient values of $v_j$ and $w_{jk}$
    \item Sparseness of the physical qubit graph
    \item Noise in the system
\end{enumerate}
We hope as the technology matures and other companies come into the market, some of these issues would get alleviated. The good news is that we can use post-processing techniques like the ones by Dorband\cite{dorband2018method,dorband2018extending} to improve the solution and increase the precision. 

\end{appendices}
\end{document}